\documentclass[fleqn,10pt]{wlscirep}
\usepackage[utf8]{inputenc}

\usepackage[T1]{fontenc}
\usepackage{amssymb}
\usepackage{graphicx}
\usepackage{color}
\usepackage{caption}
\usepackage{hyperref}

\usepackage{amsmath}
\usepackage{braket}
\usepackage{graphicx}
\usepackage{appendix}
\usepackage{subcaption}

\newcommand{\p}{q }

\title{Resilience of the topological phases	to frustration}

\author[1,2]{Vanja Mari\'{c}}
\author[1]{Fabio Franchini}
\author[1]{Domagoj Kui\'{c}}
\author[1]{Salvatore Marco Giampaolo$^*$}

\affil[1]{Division of Theoretical Physics, Rudjer Bo\v{s}kovi\'{c} Institute, Bijeni\v{c}ka cesta 54, 10000 Zagreb, Croatia}
\affil[2]{SISSA and INFN, via Bonomea 265, 34136 Trieste, Italy}

\begin{abstract}
Recently it was highlighted that one-dimensional antiferromagnetic spin models with frustrated boundary conditions, i.e. periodic boundary conditions in a ring with an odd number of elements, may show very peculiar behavior.
Indeed the presence of frustrated boundary conditions can destroy the local magnetic orders presented by the models when different boundary conditions are taken into account and induce novel phase transitions.
Motivated by these results, we analyze the effects of the introduction of frustrated boundary conditions on several models supporting (symmetry protected) topological orders, and compare our results  with the ones obtained with different boundary conditions. 
None of the topological order phases analyzed are altered by this change.
This observation leads naturally to the conjecture that topological phases of one-dimensional systems are in general not affected by topological frustration.
\end{abstract}
\begin{document}
\maketitle
\flushbottom

\begin{flushright}   
	RBI-ThPhys-2020-20
\end{flushright}

\section{Introduction}

Complex systems and their different ordered phases have always attracted a large interest, not only from a purely speculative point of view but also for the innumerable technological applications that exploit their characteristics~\cite{Baibich1988,Binasch1989,Madsen2004}.
In the middle of the last century, all the different phases of many-body systems obeying the laws of classical mechanics were classified, using the Landau theory~\cite{Landau1937}.
According to this theory, different phases are characterized by different order parameters.
Each one {of them} is uniquely associated to a particular kind of order that is related to the violation of a specific symmetry (spontaneous symmetry breaking~\cite{Anderson1997}).
It is therefore evident that, in the framework of Landau theory, the key role is played by the symmetries of the system, while other aspects, such as the boundary conditions, fall into the background and, generally, are considered not to be relevant for the presence or the absence of a particular kind of order.

When physicists began to extend their interest to the study of many-body models obeying quantum mechanics, Landau theory was immediately borrowed to study also the quantum regime~\cite{Sachdev2011}.
Notwithstanding its undoubted success, after a few years, it became clear that it was unable to catch all the aspects of the richer landscape that comes out from the different quantum complex systems.
{Indeed, nematic ordered phases~\cite{Shannon2006} show a situation in which the violation of the same symmetry is associated to different and non-equivalent order parameters, depending on the model under analysis~\cite{Lacroix2011}, while topologically ordered phases~\cite{Wen1989,Wen1990} do not admit local order parameters, but rather non-local ones.}
This implies that the classification of different ordered phases in terms of the {local} symmetry violated by the order parameter, which is the cornerstone of Landau theory, is not able to provide a complete classification of the quantum phases of matter.

Moreover, in recent years, even the assumption that some aspects of complex systems, such as boundary conditions, do not affect the onset of an ordered phase has been questioned.
{Indeed, it was observed that} particular boundary conditions, named frustrated boundary conditions (FBCs), i.e. periodic boundary conditions with an odd number of lattice sites, when paired with antiferromagnetic interactions, {induce a topological frustration that can change several properties of the system such as the energy gap and the entanglement entropy~\cite{Dong2016, Giampaolo2018}. 
Moreover the presence of the topological frustration can affect the exponential/algebraic dependence of the two-points correlation functions on the distance~\cite{Dong2016,mari2019frustration}, and prevent the formation of the AFM ordered phase that characterizes the same model without frustration~\cite{mari2019frustration,MaricToeplitz}.
Depending on the symmetry of the model, the frustration can give birth to a first-order phase transition that separates a mesoscopic ferromagnetic phase from a peculiar antiferromagnetic order, with a magnetization slowly varying in space and incommensurate with the lattice length~\cite{mari2020frustration, Torre2020}.
In other models, the effects of the topological frustration can be even stronger coming to the point of preventing the creation of any local order~\cite{MariNoOrder2021} and thus inducing a change in the nature of the macroscopic phases, and related phase transitions, in the system~\cite{mari2021Nematic}.}

Behind this violation of one of the main prescriptions of Landau theory is that FBCs induce a geometrical frustration in the system~\cite{Toulouse1977,Vannimenus1977}.
Frustration is an extremely broad phenomenon, that can have a quantum and/or a geometrical/topological origin with a huge spectrum of effects~\cite{Sadoc1999,Lacroix2011,Diep2013}.
However, at its deep core, frustration always comes out from the impossibility to satisfy simultaneously all local constraints in a many--body system~\cite{Wolf2003,Giampaolo2011,Marzolino2013}.
The usual way to understand frustration is to think about a classical system made of dichotomous objects, i.e. a classical equivalent of quantum spins that can assume only two values, namely $\pm 1$, whose behavior is governed by a Hamiltonian with a two--body short--range Ising--like antiferromagnetic interactions.
While, locally, there is no problem to minimize an antiferromagnetic Ising bond, when the dichotomous objects are arranged in loops made by odd numbers of sites, at least one of such bonds needs to display a ferromagnetic alignment.
Despite its simplicity, this example captures all the main aspects of frustration.
Indeed, even in more complex lattices, the presence of geometrical frustration can be traced back to the existence of frustrated loops made by an odd number of antiferromagnetic bonds~\cite{Toulouse1977}.

While some phenomenology on the effects of geometrical frustration on antiferromagnetic phases in quantum many-body has been accumulated, nothing has still been done for phases exhibiting topological orders. 
In the present paper, our goal is to fill this gap.
Therefore, we will focus on several exactly solvable models known to exhibit a weak form of topological order, namely a Symmetry Protected Topological Order (SPT), which can be broken by a perturbation not respecting the symmetry of the Hamiltonian without closing the gap~\cite{PhysRevB.85.075125,PhysRevB.83.035107,PhysRevB.96.165124,doi:10.1146/annurev-conmatphys-031214-014740,zeng2018quantum}, {and we analyze the effect of imposing frustrated boundary conditions on them. }



At first, in Sec.~\ref{sec Cluster}, we study the so-called Cluster--Ising model~\cite{Smacchia2011,Giampaolo2015} that is a one--dimensional model, in which a three--body cluster interaction competes with an antiferromagnetic Ising one.
Despite its apparent complexity, the model admits an analytical solution obtained by mapping the spins degrees of freedom into spinless fermions using the Jordan-Wigner transformation~\cite{Jordan1928}.
Since there is an interplay between two different types of interactions, the model presents a transition between a phase dominated by the cluster interaction and one in which the leading term is the Ising one.
While in the second region the system (usually) admits a magnetically ordered phase, when the cluster interactions dominate over the antiferromagnetic ones, the model is known to exhibit a symmetry protected topological order.
{We will show that, while the magnetic phase results to be deeply affected by the rising of topological frustration, the SPT ordered phase is insensitive to the particular choice of boundary conditions, a fact that we explain considering simple properties of the two and three body interactions. 
The fact that the sensitivity of the model to the change of boundary conditions is a function of the macroscopic phase and not only of the model considered, strongly suggests that such a property must extend to all other models that present SPT ordered phases.} 
%

{To test such a suggestion we turn, in Sec.~\ref{sec AKLT}, to the AKLT chain~\cite{Affleck1987,Affleck1988}, which is a one-dimensional model, with a $SO(3)$ symmetric Hamiltonian, describing spin-$1$ degrees of freedom interacting antiferromagnetically. The model supports a unique ground state (up to boundary state degeneracies), separated from the rest of the eigenstates by a finite gap, and it exhibits exponentially decaying correlation functions. It is characterized by spins paired into valence-bonds and symmetry protected topological order~\cite{PhysRevB.96.165124, zeng2018quantum}.}
The model with periodic boundary conditions, including FBCs, has been already studied in details~\cite{Affleck1988,Tasaki2020} so we simply summarize and discuss the results relevant for this work showing that also in this model {frustrated boundary conditions do not affect the model and its order.}

{Finally, in Sec.~\ref{sec Kitaev} we consider last of the 1D topological models in our survey, namely the Kitaev chain~\cite{Kitaev_2001}. 
With open boundary conditions, the Kitaev chain is known to be exactly mappable} to the quantum XY chain in a transverse magnetic field~\cite{McCoy2,Franchini:2016cxs}, but moving to periodic BC there are some subtleties, that play an important role when we wish to analyze the effect of frustration.
In all cases, the Kitaev chain can be reduced to a free fermionic problem, and therefore diagonalized analytically~\cite{LIEB1961407,Franchini:2016cxs}. {Exploiting this approach we highlight that, regardless on the macroscopic phase, the Kitaev chain is completely unaltered by moving from an even to an odd number of sites (or viceversa).}

Thus in all models analyzed we find that their symmetry protected topological orders are {either resilient to the topological frustration imposed by frustrated boundary conditions, or that the system is such that frustrated boundary conditions do not introduce any incompatibility in the simultaneous minimization of the local and global structure of the Hamiltonian. 
These observations} naturally drives us to conjecture that SPT ordered phases (and, a fortiori, topological ordered phases) {in one dimension are not affected by frustrated boundary conditions.}
This result could be, to some extent, counter intuitive.
In fact, it shows that topological phases are not affected by the (real-space) topology of the system.
Most of all, {our conjecture} places a further clear distinction between phases characterized by non-global order parameters, as the magnetic and the nematic ones, and topologically ordered ones, {at least in one dimension}. 

\section{The Cluster-Ising model}\label{sec Cluster}

\subsection{The Model}\label{sec model}

Let us start by introducing the Cluster-Ising model.
We consider a system made of spins-$\frac{1}{2}$, in which a two-body antiferromagnetic Ising pairing competes with a three-body cluster interaction. The Hamiltonian of such model reads
\begin{equation}
\label{Hamiltonian_1}
H=\cos\phi \sum\limits_{j=1}^{N} \sigma_j^x\sigma_{j+1}^x +\sin\phi\sum\limits_{j=1}^{N} \sigma_{j-1}^y \sigma_{j}^z \sigma_{j+1}^y ,
\end{equation}
where $\sigma_j^\alpha$, for $\alpha=x,y,z$, are the Pauli operators on the $j$-th spin and the parameter $\phi$ allows to change the relative weight between the two terms. Sine our goal is to study the effect of topological frustration, we assume periodic boundary conditions $\sigma_{j+N}^\alpha = \sigma_j^\alpha$ and that $N$ is an odd number ($N=2M+1$).
Due to the existence of an analytical solution, the family of spin-$1/2$ cluster models was intensively studied in the past years~\cite{Smacchia2011,Pachos2004,Montes2011,Giampaolo2015,GiampaoloH2015,Zonzo2018}. For the Cluster--Ising model in eq.~\eqref{Hamiltonian_1} it is well-known that for {$\phi\in(-\pi/4,\pi/4)$ the model is in an antiferromagnetic Ising phase, for $\phi\in(3\pi/4,5\pi/4)$ in a ferromagnetic phase, while for} $\phi\in(\pi/4,3\pi/4)$ and $\phi\in(-3\pi/4,-\pi/4)$, where the many-body interactions dominate over the Ising ones, the model exhibits a symmetry protected topological order.
Such order can be characterized by a non-zero expectation value for the non-local string operator, defined as
\begin{equation}\label{string}
O(r)=\sigma_1^y\sigma_2^x \Big(\bigotimes_{j=3}^{r}\sigma_j^z \Big) \sigma_{r+1}^x\sigma_{r+2}^y.
\end{equation}

Before we start the detailed solution of the model based on the Jordan-Wigner transformation, let us make some general considerations.
Both from the seminal Toulouse's works~\cite{Toulouse1977,Vannimenus1977} for classical models and from their generalization to the quantum regime~\cite{Giampaolo2011,Marzolino2013}, we have that, to determine whether or not a model is geometrically or topologically frustrated, we need two elements.
The first of these elements is a prototype model, i.e. a model in which frustration is absent.
The second element is a set of local unitary operators, by which we try to reduce the model under analysis to the prototype one.
If it is possible to find such a set of local unitary operations that map our Ising--like model into the prototype one, then the system is free from geometrical frustration. Otherwise, we have that the system is geometrically or topologically frustrated.

In the case of an antiferromagnetic Ising model the prototype model is the model in which each bond is turned into a ferromagnetic one.
Let us now focus on the case of a one--dimensional lattice with periodic boundary conditions in which all bonds between neighboring spins are antiferromagnetic. Such model is exactly the Ising term of the Hamiltonian in eq.~\eqref{Hamiltonian_1} and can be written as $\sum_{i=1}^N\sigma_i^x\sigma_{i+1}^x$. If $N$ is even, the sign of all the terms $\sigma_j^x\sigma_{j+1}^x$ can be inverted simply by inverting every second spin, starting from $j=1$, hence reducing it to a purely ferromagnetic Ising model and thus proving that the model is not frustrated.
On the contrary, with odd $N$, {it is impossible to bipartite the system in two sets made by the same number of elements making it impossible to find a suitable transformation, acting on only one of the two subsets, that transforms it into a purely ferromagnetic model}, hence proving the presence of frustration.

Let us now consider the cluster term of the eq.~\eqref{Hamiltonian_1}, i.e. $\sum_{j}\sigma_{j-1}^y\sigma_{j}^z\sigma_{j+1}^y$.
Differently from the Ising case, inverting every single spin of the lattice through a unitary operation generated by the spin operators $\sigma_j^y$ it is always possible to change the global sign of the model.
{As a consequence}, the three-body cluster interaction is not expected to show any geometrical frustration. This should be contrasted with the case of cluster interaction with an even number of sites, such as the 2-Cluster-Ising chain~\cite{GiampaoloH2015,Zonzo2018}, which do not have topological order. According to the argument above, in these cases we expect that frustration will be effective and indeed we {have} proven that to be the case~\cite{MariNoOrder2021,mari2021Nematic}.

When we have the simultaneous presence of both the interactions, arguments like those just made become more complex.
However, in the parameter region in which the Ising type interaction dominates over the cluster one, taking inspiration from arguments such as adiabatic deformation~\cite{Hastings2005}, we expect the system to present signatures of the presence of geometrical frustration.
Viceversa, in the region dominated by cluster-type interaction, these signatures can be expected to be absent.
Such region is expected to show a symmetry protected topological order~\cite{Smacchia2011}, and therefore it is expected to be unaffected by the presence of frustration.

Let us now turn to the exact solution of the Cluster--Ising model in presence of frustrated boundary conditions.
It is well known~\cite{Smacchia2011,GiampaoloH2015} that it can be diagonalized exactly, using the Jordan--Wigner transformation
\begin{equation}\label{JW}
c_j=\Big(\bigotimes_{l=1}^{j-1}\sigma_l^z\Big)\frac{\sigma_j^x+\imath\sigma_j^y}{2}, \quad c_j^\dagger=\Big(\bigotimes_{l=1}^{j-1}\sigma_l^z\Big)\frac{\sigma_j^x-\imath\sigma_j^y}{2},
\end{equation}
that maps spins into spinless fermions.
In the process of diagonalization, details can be found in the Supplementary Information, the Hamiltonian is divided in the two parity sectors of $\Pi^z=\bigotimes_{j=1}^N\sigma_j^z$,
\begin{equation}
\label{Hamiltonian_2}
H=\frac{1+\Pi_z}{2}H^+ \frac{1+\Pi_z}{2} + \frac{1-\Pi_z}{2}H^- \frac{1-\Pi_z}{2} \; ,
\end{equation}
and in each sector the Hamiltonian can be written in terms of free fermionic operators
\begin{equation}
\label{Hamiltonian_3}
H^\pm=\sum\limits_{q\in\Gamma^\pm}^{} \varepsilon_q \left(a_q^\dagger a_q-\frac{1}{2}\right) ,
\end{equation}
where $a_q$ are Bogoliubov fermions.
The fermionic momenta $q$ in eq.~\eqref{Hamiltonian_3} belong to two different sets, respectively $q \in \Gamma^+ = \{\frac{2\pi}{N}(k+\frac{1}{2})\} $ for the even parity sector ($\Pi^z=1$) and $q \in \Gamma^-= \{\frac{2\pi}{N}k\} $ for the odd one ($\Pi^z=-1$), where, in both cases, $k$ runs over all integers between 0 and $N-1$.

To each fermionic momentum is associated an energy, given by
\begin{eqnarray}
\label{Energy_mode} 
\varepsilon_q   &  =2\sqrt{1 +\sin 2\phi \cos 3q } & \quad \forall \,q\neq 0,\pi,\nonumber \\
\varepsilon_0  & =2( \sin\phi+\cos\phi)& \quad  q=0\in \Gamma^- , \\
\varepsilon_\pi & = 2(\sin\phi-\cos\phi) \nonumber & \quad q=\pi\in \Gamma^+.
\end{eqnarray}
It is worth noting that the momenta $0\in \Gamma^-$ and $\pi\in\Gamma^+$ (since we study the case in which $N$ is odd), are different from the others because: a) they do not have a corresponding opposite momentum; b) their energies can be negative.

From eqs.~(\ref{Energy_mode}) it is easy to determine the ground states of the system starting from the vacuum of Bogoliubov fermions in the two sectors ($\ket{0^\pm}$), which, by construction, have positive parity $\Pi^z=1$, and taking into account the modes with negative energy and the parity requirements. {We are going to examine the antiferromagnetic phase and the cluster phase separately. To compute the ground state expectation of observables it is going to be convenient to use} the Majorana fermions, defined as
\begin{equation}\label{Majorana Dirac}
A_j=c_j^\dagger+c_j, \quad B_j=\imath(c_j^\dagger-c_j),
\end{equation}
which are related to the spin operators as
\begin{equation}
A_j= \Big(\bigotimes\limits_{l=1}^{j-1} \sigma_l^z \Big) \sigma_j^x \; , \quad B_j= \Big(\bigotimes\limits_{l=1}^{j-1} \sigma_l^z \Big) \sigma_j^y.
\end{equation}
{On the basis of Wick theorem the expectation values of observables are determined by the two-point correlators of Majorana fermions.}

\subsubsection{Antiferromagnetic phase}
\label{sec:AFMISING}

{
In studying the antiferromagnetic phase $\phi\in(-\pi/4,\pi/4)$ we focus on the parameter region $\phi\in(-\pi/4,0)$
without loosing generality since to any ground state $\ket{g(\phi)}$ corresponds the ground state $\ket{g(-\phi)}=\Pi^x\ket{g(\phi)}$ 
In particular, the spin-correlations functions are the same in the two states, $\bra{g(-\phi)}\sigma_1^x\sigma_{1+r}^x\ket{g(-\phi)}=\bra{g(\phi)}\sigma_1^x\sigma_{1+r}^x\ket{g(\phi)}$, and, similarly, the magnetization $\bra{g(\phi)}\sigma_j^x\ket{g(\phi)}=\bra{g(-\phi)}\sigma_j^x\ket{g(-\phi)}$.
}

{
We find that the ground state degeneracy depends on whether (odd) $N$ is divisible by $3$. 
When $N$ is not divisible by $3$ the ground state is single, given by $\ket{g}=a_0^\dagger\ket{0^-}$, corresponding to the energy minimizing mode $q=0$. 
When $N$ is divisible by $3$ there are three energy minimizing modes, given by $q=0,2\pi/3,-2\pi/3$. The ground state manifold is thus three-fold degenerate and a general ground state is a superposition
\begin{equation}\label{gs three fold}
\ket{g}=(u_1 a_0^\dagger+u_2 a_{\frac{2\pi}{3}}^\dagger+u_3 a_{-\frac{2\pi}{3}}^\dagger)\ket{0^-},
\end{equation}
where the normalization $\sum_{j=1}^3|u_j|^2=1$ is assumed. 
However, regardless the dimension of the ground state manifold, it always falls into a single $\Pi^z$ sector, with an energy gap above it that closes as $1/N^2$ for large (odd) $N$. 
}

{
The Majorana correlators in the ground state \eqref{gs three fold} are found to be
\begin{equation}
\begin{split}\label{AA superposition correlator}
&\braket{A_jA_l}_g=\delta_{jl}-\frac{2\imath}{N}(|u_2|^2-|u_3|^2)  \sin\Big[\frac{2\pi}{3}(j-l)\Big]-\frac{2\imath}{N}\Big[(u_1^*u_2-u_3^*u_1)e^{\imath\frac{\pi}{3}(j+l-1)} +\textrm{c.c.}\Big] \sin\Big[\frac{\pi}{3}(j-l)\Big],\\
&\braket{B_jB_l}_g=\delta_{jl}-\frac{2\imath}{N}(|u_2|^2-|u_3|^2)  \sin\Big[\frac{2\pi}{3}(j-l)\Big]-\frac{2\imath}{N}\Big[(u_1^*u_2-u_3^*u_1)e^{\imath\frac{\pi}{3}(j+l+1)} +\textrm{c.c.}\Big] \sin\Big[\frac{\pi}{3}(j-l)\Big],\\
-\imath & \braket{A_jB_l}_g \stackrel{N\to\infty}{\simeq}\int_{0}^{2\pi} \frac{\cos \phi + \sin \phi \ e^{-\imath 3q}}{|\cos \phi + \sin\phi\ e^{-\imath 3q}|} e^{-\imath\p(j-l-1)}\frac{dq}{2\pi} -\frac{2}{N}\Big\{ |u_1|^2+(|u_2|^2+|u_3|^2)\cos\Big[\frac{2\pi}{3}(j-l-1)\Big] \Big\} \\
&-\frac{2}{N}\Big[(u_1^*u_2+u_3^*u_1) e^{\imath\frac{\pi}{3}(j+l)} +\textrm{c.c.} \Big]\cos\Big[\frac{\pi}{3}(j-l-1)\Big] - \frac{2}{N}\Big[u_2^*u_3 \ e^{-\imath\frac{2\pi}{3} (j+l)} +\textrm{c.c.}  \Big]\,,
\end{split}
\end{equation}
while, for $N$ not divisible by $3$, the correlators can be reproduced from the previous expressions by taking formally $u_2=u_3=0$. The exact finite-$N$ result for the last correlator can be found in the Supplementary Information.
In the correlators we can see corrections of order $1/N$, that would not be present in a model without topological frustration. 
Although they vanish in the thermodynamic limit, they cannot be neglected, as they can influence the spin-correlation functions at large distances, i.e. for a distance that scales with the dimension of the system. 
Indeed, as shown in the Supplementary Information, we find the spin-correlation functions
\begin{equation}\label{spin correlations n 1}
\braket{\sigma_1^x\sigma_{1+r}^x}_{g}\stackrel{r\to\infty}{\simeq}(-1)^r(1-\tan^2\phi)^{3/4} \bigg(1-\frac{2r}{N}\bigg).
\end{equation}
The correlations at large distances $r$ decay linearly and for the most distant spins on the ring, separated by $r\approx N/2$, they vanish in the thermodynamic limit.
Coherently also the magnetization order parameter vanishes in the presence of FBCs and antiferromagnetic interactions. 
Thus, topological frustration affects the antiferromagnetic phase of the Cluster-Ising chain, closing the energy gap and destroying the magnetization and spin-correlations at large distances, similarly to several other frustrated chains~\cite{Dong2016,Giampaolo2018,mari2019frustration}.}

\subsubsection{Cluster phase}

We find that for $\phi\in(\pi/4,3\pi/4)$ the ground state is $\ket{g}=\ket{0^+}$
and it is separated by a finite energy gap from the excited states above it.
Similarly, for $\phi\in(-3\pi/4,\pi/4)$ the ground state is $\ket{g}=a_0^\dagger\ket{0^-}$, also with a finite energy gap above it.
These two regions of the model's phase space are those known to display symmetry protected topological order \cite{Smacchia2011,GiampaoloH2015}.

As shown in the Supplementary Information, the correlators of Majorana fermions are $\braket{A_jA_l}_g=\braket{B_jB_l}_g=\delta_{jl}$ and
\begin{equation}
\braket{A_jB_l}_g\!\!\!\stackrel{N\to\infty}{\simeq}\!\!\!\imath\int\limits_{0}^{2\pi} \frac{\sin\phi+\cos\phi \ e^{i3\p}}{|\sin\phi+\cos\phi \ e^{i3\p}|} e^{-iq(j-l+2)} \frac{dq}{2\pi}, \label{Majorana correlators}
\end{equation}
which is the same result as without frustration~\cite{Smacchia2011,GiampaoloH2015}, {i.e. the same result would be obtained for even $N$. There is no corrections of order $1/N$ as in the antiferromagnetic phase}.

The consequence of this result is that {the expectation value of any bulk observable remains the same with frustrated boundary conditions, generic periodic BC or, arguably, open BC}. Indeed, in the {antiferromagnetic phase} the effect of frustration arises as a correction to the Majorana correlation function.
Since, any observable can be expressed in terms of Pauli spin-operators, while Pauli spin-operators can be expressed as a product of Majorana fermions, the expectation value of any observable can be expressed as an expectation of a product of Majorana fermions, which is by Wick theorem determined by two-point correlators of Majorana fermions.
Therefore, since the two-point correlators of Majorana fermions are the same as without frustration for large $N$, and since the same applies for the Jordan-Wigner transformation \eqref{Majorana Dirac}, so is the expectation value of any bulk observable. 


In particular, in the Supplementary Information we compute the expectation value of the string operator.
We obtain
\begin{equation}\label{string order}
\!\!\!\!\braket{O(r)}_g\!\!\stackrel{r\to\infty}{\simeq}\!\! \begin{cases}
(1-\cot^2\phi)^{\frac{3}{4}} , & \phi\in(\frac{\pi}{4},\frac{3\pi}{4})\\
(-1)^r (1-\cot^2\phi)^{\frac{3}{4}} , & \phi\in(-\frac{3\pi}{4},-\frac{\pi}{4})
\end{cases}
\end{equation}
as without frustration \cite{GiampaoloH2015}.
On the contrary, in the topologically ordered phases, the expectation values of the operators $\braket{\sigma_j^\alpha}$ for $\alpha=x,y,z$ are zero.
Namely, since the ground state does not break the parity symmetry $\Pi^z$ of the model we have immediately $\braket{\sigma_j^x}=0,\ \braket{\sigma_j^y}=0$, while the relation $\braket{\sigma_j^z}=0$ follows from the equality $\sigma_j^z=-\imath A_j B_j$ and the property that the corresponding integral in \eqref{Majorana correlators} vanishes.

The non-zero expectation value of the non-local string operator, and zero expectation value of local observables, such as spin operators, characterizes the symmetry protected topological ordered phase in the Cluster-Ising model. It is, hence, proved that such a phase is not affected by topological frustration.

\section{AKLT model}\label{sec AKLT}

The AKLT model \cite{Affleck1987,Affleck1988} is a one dimensional model, describing spin-$1$ degrees of freedom interacting antiferromagnetically. It is defined by the $SO(3)$ symmetric Hamiltonian
\begin{equation}
	H=\sum_{j=1}^N \Big[\vec{S}_j\cdot\vec{S}_{j+1}+\frac{1}{3}\big(\vec{S}_j\cdot\vec{S}_{j+1}\big)^2 \Big].
	\label{AKLTHam}
\end{equation}
The FBCs are achieved by imposing an odd number of lattice sites $N=2M+1$ and periodic BC $\vec{S}_{j+N}=\vec{S}_j$. 
{Naively, we can expect that FBCs induce frustration in the AKLT model, by considering its classical limit, and thus neglecting the second term in eq.~\eqref{AKLTHam}. With periodic boundary conditions and even $N$ the energy is clearly minimized by the perfectly staggered configuration $S_j=(-1)^j$ or its flipped version. However, for odd $N$, corresponding to FBCs, such a configuration is impossible because the lattice is not bipartite. Therefore, not all local terms in the Hamiltonian can be minimized simultaneously, and the Hamiltonian is frustrated.}

{Turning back to the quantum case, the AKLT Hamiltonian in eq.~\eqref{AKLTHam} acquires a (symmetry protected) topological order and it is thus an ideal candidate to test our conjecture that this property protects it from the effect of FBCs. It is easy to check that  eq.~\eqref{AKLTHam}} can be written as a sum of projectors
\begin{equation}
	H = 2 \sum_{j=1}^N \left[ P^{(2)} (\vec{S}_j,\vec{S}_{j+1}) - \frac{1}{3} \right] ,
\end{equation}
where $P^{(2)} (\vec{S}_j,\vec{S}_l)$ projects the state of two spin-$1$ at sites $j$ and $l$ into their spin-$2$ representation.

The AKLT model with both open and periodic BC has been studied in detail in \cite{Affleck1988} (for a more pedagogical approach see the book \cite{Tasaki2020}). It is known that the ground state is unique with PBC, and four-fold degenerate with open BC, with this degeneracy related to the existence of edge states and thus not influencing the expectation values of bulk observables, which are the same in the different ground states (similarly to what happens in the Cluster-Ising model and the Kitaev chain that we will analyze later).
The AKLT ground state, with periodic boundary conditions, is a valence-bond state. To show this, one represents each spin-$1$ degree of freedom through two spin-$1/2$ in their triplet representation. Then, these spin-$1/2$ on neighboring sites are paired as singlet to prevent the ferromagnetic alignment penalized by the Hamiltonian.  Denoting by $|\alpha\rangle_j$, $|\beta \rangle_j$ the two spin-$1/2$ on the $j$-th site, the valence bond state is
\begin{equation}
	| V \rangle_j \equiv {\bf V}^{\beta_j,\alpha_{j+1}} |\beta_j \rangle_j |\alpha_{j+1} \rangle_{j+1} = 
	\frac{1}{\sqrt{2}} \Big( | \uparrow \rangle_j  | \downarrow \rangle_{j+1} - 
	| \downarrow \rangle_j  | \uparrow \rangle_{j+1} \Big) \; ,
\end{equation}
where ${\bf V} \equiv \frac{1}{\sqrt{2}} \left( \begin{matrix} 0 & 1 \\ -1 & 0 \\ \end{matrix} \right)$. 
We then construct the ground state of \eqref{AKLTHam} as
\begin{equation}
	|GS\rangle \equiv \prod_{l=1}^N {\hat P}^{(1)}_l 
	\bigotimes_{j=1}^N | V \rangle_j \; ,
\end{equation}
where 
\begin{equation}
	{\hat P}^{(1)}_l \equiv {\bf P}^{\sigma}_{\alpha,\beta} | \sigma \rangle_j
	\langle \alpha | _j \langle \beta | =
	| +1 \rangle_j  \langle \uparrow \uparrow | +
	\frac{1}{\sqrt{2}} | 0 \rangle_j \Big( \langle \uparrow \downarrow | + 
	\langle \downarrow \uparrow | \Big) + 
	| -1 \rangle_j  \langle \downarrow \downarrow | \; ,
\end{equation}
projects the two spin-$1/2$ into their spin-$1$ representation, with the Clebsch-Gordan coefficients ${\bf P}^{+1} \equiv \left( \begin{matrix} 1 & 0 \\ 0 & 0 \\ \end{matrix} \right)$, ${\bf P}^{0} \equiv \frac{1}{\sqrt{2}} \left( \begin{matrix} 0 & 1 \\ 1 & 0 \\ \end{matrix} \right)$, and ${\bf P}^{-1} \equiv \left( \begin{matrix} 0 & 0 \\ 0 & 1 \\ \end{matrix} \right)$, and where the $N+1$ site is identified with the first, since periodic boundary conditions are assumed. As already discussed in \cite{Affleck1988}, the only difference between having even or odd $N$ is the need of a different index contraction in the spinorial representation of the valence bond state. However, {the even-odd choice does} not change the gapped nature of the system and the bulk behavior of the correlation functions \cite{Affleck1988,Tasaki2020}.

In particular, the string order parameters that encode the topological order of the AKLT model 
\begin{eqnarray}
	S_{j,l}^{(\alpha)} & \equiv & \langle GS | S_j^\alpha e^{\imath \pi \sum_{n=j+1}^{l-1} S_n^\alpha} S_l^\alpha | GS \rangle / \langle GS | GS \rangle \; ,
	\\
	S^{(\alpha)} & \equiv & \lim_{l-j \to \infty} \lim_{N \to \infty} S_{j,l}^{(\alpha)}=\frac{4}{9} \; ,
\end{eqnarray}
for $\alpha=x,y,z$, remain non-zero and unchanged.


We thus conclude that the (symmetry protected) topological order of the AKLT model is not affected by {frustrated boundary conditions}.

\section{Kitaev chain}\label{sec Kitaev}

The Kitaev chain~\cite{Kitaev_2001} is a model of a spinless fermions topological superconductor for which the Hamiltonian reads
\begin{equation}\label{Kitaev Hamiltonian}
H=
- \mu \sum_{j=1}^{N} \Big(c_j^\dagger c_j-\frac{1}{2}\Big)
- \sum_{j=1}^{N} \left[
w \: (c_j^\dagger c_{j+1}+\textrm{h.c.})
-\Delta \: (c_jc_{j+1}+\textrm{h.c.}) \right],
\end{equation}
where $\mu$ is the chemical potential, $w$ is the hopping amplitude and $\Delta$ is the superconducting gap.
As we have done so far, also with the Kitaev model we will focus on the case of FBCs, i.e. with periodic boundary conditions ($c_{j+N}=c_j$
) and an odd number of lattice sites $N=2M+1$.
However, before we start the analysis of the case with periodic boundary conditions, let us summarize the main results that were obtained for the open ones.

It is well-known that the Kitaev chain with open boundary conditions, that can be obtained from eq.~\eqref{Kitaev Hamiltonian} restricting the range of the second sum up to $j=N-1$, can be mapped, inverting the Jordan-Wigner transformation in eq.~\eqref{JW}, to the quantum XY chain in transverse field~\cite{McCoy1968,Franchini:2016cxs}
\begin{equation}
H=-\sum_{j=1}^{N-1} \left[ 
\frac{w+\Delta}{2} \: \sigma_j^x\sigma_{j+1}^x
+\frac{w-\Delta}{2} \: \sigma_j^y\sigma_{j+1}^y \right] +\frac{\mu}{2}\sum\limits_{j=1}^N \sigma_j^z.
\end{equation}
Hence, exactly as the correspondent spin model, in the thermodynamic limit, it shows a phase transitions at $\mu=\pm 2w$.
Such quantum phase transition separates a topologically trivial phase for $|\mu|>2|w|$ without edge modes in the open chain, from a topologically ordered phase $|\mu|<2|w|$ characterized by the presence of Majorana edge modes. {The latter, for the spin chain, was shown to be affected by FBCs when the Ising term promotes an AFM order, with a phenomenology similar to that of the Cluster-Ising chain we discussed in Sec.~\ref{sec:AFMISING}. Due to the relation between the XY and the Kitaev chain, it is natural to wonder if the latter can also be affected by FBCs, although we cannot compare the spinless model to a related classical one which is affected by frustration.}

Moving from open to periodic boundary conditions, the exact equivalence between the fermionic and the spin model ceases to exist.
The reason for such quite surprising result is connected to the fact that the Jordan-Wigner transformations breaks the invariance under spatial translation, by selecting a reference site.
This implies that the interactions terms between the first and the last spin of a chain are no more mapped in a standard two--body fermionic term, but in a string term that makes it impossible to map a short-range fermionic model into a short range spin model.
To provide an example, we have that the term $\sigma_N^x\sigma_{1}^x$ is mapped into the string operator $-\Pi^z(c_N^\dagger-c_N)(c_1^\dagger+c_1)$ where $\Pi^z$ is the parity operator along the $z$ axis that has support on the whole lattice.
A similar result stands also for $\sigma_N^y\sigma_1^y$.
When either $w \pm \Delta<0$ (and  $|\mu|<2|w|$), the XY chain, {with an odd number of sites,} becomes frustrated: the energy gap above the ground states closes (algebraically) in the thermodynamic limit \cite{Dong2016}, the correlation functions acquire peculiar algebraic corrections and the entanglement entropy violates the area law \cite{Giampaolo2018}, and the AFM local order is replaced by either a ferromagnetic mesoscopic order of by a AFM incommensurate modulated one (with a quantum phase transition separating the two) \cite{mari2019frustration,mari2020frustration}. 

From a physical perspective, the main difference between the Kitaev and the XY chain is that only the former supports (symmetry protected) topological order. 
Given the close relation between the two models and the fact that the effects of frustration have already been established for the spin chain, {we may wonder whether there is an even-odd effect also in the Kitaev chain}. 

As presented in the Supplementary Information, exploiting the approach illustrated in Ref.~\cite{Franchini:2016cxs}, we can diagonalize the Kitaev chain with periodic boundary conditions and odd $N$ obtaining
\begin{equation}
H=\sum_{q\in\Gamma}\varepsilon_q \Big(a_q^\dagger a_q-\frac{1}{2}\Big),
\end{equation}
where $a_q$ are Bogoliubov fermions, whose momenta belong to the set $\Gamma=\{\frac{2\pi}{N}k\}$, with $k$ running over integers between $0$ and $N-1$.
It is worth to note that, assuming $N$ to be odd, $k=\pi$ is not an allowed momentum in this model.
The dispersion relation is given by
\begin{align}
& \varepsilon_q=\sqrt{(4w\cos q+\mu)^2+4\Delta^2\sin^2 q} , \quad q\neq0\\
& \varepsilon_0=-2w-\mu,
\end{align}
Similarly to the Cluster-Ising model case, the mode $q=0$ is special because it is the only one in which the energies can be negative.
The eigenstates of the model are constructed by populating the vacuum state $\ket{0}$.
Taking into account the dispersion relation, it is easy to see that the ground state of the Kitaev chain with periodic frustrated boundary conditions, is always non-degenerate, with a finite energy gap above it, except at the phase transition points $\mu=\pm 2w$, where the spectrum becomes gapless and relativistic.
Similarly to the Cluster-Ising model, in the topologically ordered phase $|\mu|<2|w|$, the ground state degeneracy, in the thermodynamic limit, is different from {the one} that characterizes the model with open boundary conditions. While the former is two-fold, with PBCs the ground state is unique.
Nevertheless, the expectation values of bulk observables are the same, as we now show.

For this purpose, we define Majorana fermions using eq.~\eqref{Majorana Dirac}. Then, similarly to the Cluster-Ising model, all operators acting on the Fock space generated by $c_j^\dagger$ can be expressed in terms of Majorana fermions, and using the Wick theorem it follows that the ground state expectation value of any observable is determined by the two-point correlators of Majorana fermions.
As shown in the Supplementary Information, the two-point correlators of Majorana fermions in the ground state, in all parameter regions of the model, are $\braket{A_jA_l}_g=\braket{B_jB_l}_g=\delta_{jl}$ and
\begin{equation}
\!\!\!\!
\braket{A_jB_l}_g \!\!\!\! \stackrel{N\to\infty}{\simeq} \!\!\!\! \imath \int\limits_{0}^{2\pi} \!\!\frac{2w\cos q+\mu+2\Delta\imath\sin q}{|2w\cos q+\mu+2\Delta\imath\sin q|}e^{-\imath q(j-l)}\frac{dq}{2\pi}
\end{equation}
Hence, for large $N$, the expression of the Majorana correlation functions {do} not depend on the boundary conditions {(the result for free boundary conditions follows from the equivalent spin chain\cite{McCoy1968})}.
Therefore, the expectation values of all bulk observables in the Kitaev chain with FBCs are equal to those in other settings and no difference emerges when $w\pm \Delta<0$. We conclude, in particular, that topological order in the Kitaev chain is, as in the Cluster-Ising model, not affected by {
frustrated boundary conditions}.

\section{Conclusions}\label{sec conclusions}

We have presented an analysis of the effects of {frustrated boundary conditions} on different (symmetry protected) topologically ordered phases characterizing one-dimensional models.
At first, we have focused on the one-dimensional Cluster-Ising model with an odd number of spins and periodic boundary conditions.
We presented general arguments by which the symmetry protected topological order of the cluster phase is not expected to be affected by frustration, while the antiferromagnetic phase is.
These speculative arguments { have been confirmed by our analytic results.
While in the antiferromagnetic phase FBCs close the energy gap and destroy both the spin-correlations at large distances and the magnetization, the string order parameter in the cluster phase is not affected by FBCs.}
The property that the effects of topological frustration are lost when the cluster interactions start dominating over the antiferromagnetic ones is, in some extent, similar to the situation in the frustrated Ising model~\cite{Dong2016,Giampaolo2018}, where the effects of frustration are suppressed by increasing the magnetic field, resulting in the resilience of the paramagnetic phase to geometrical frustration.

Our results on the Cluster-Ising model are even more interesting if we observe that we can directly transfer our general arguments to the $m$-cluster Ising model, studied in Refs.~\cite{GiampaoloH2015,Zonzo2018}, which consists of $m$-body cluster interaction competing with the antiferromagnetic Ising pairing. It is known that for any odd $m$, the model is characterized by a symmetry protected topologically ordered phase, which is, by our general arguments, expected not to be affected by geometrical frustration.
A question that arises naturally already at this point is whether {FBCs do not affect only} models with cluster interactions or {it} is a general property of systems supporting topological order.
For this reason, we have extended our analysis to two additional one-dimensional models that also exhibit SPT, such as the AKLT model and the Kitaev chain. In both cases, the topological phase is unaffected by {FBCs, as is also shown} by the fact that a typical effect of frustration is the closing of the energy gap \cite{mari2019frustration,mari2020frustration,Giampaolo2018,Dong2016}, which remains open in {the topological phases of} all the models analyzed.
Since we have found in various models that the topologically ordered phases analyzed so far are not affected by {FBCs}, we arrive naturally to the conjecture: {Frustrated boundary conditions do not affect symmetry protected topological phases of one dimensional systems. } However, whether the conjecture is correct or not it remains to be established for certain.
One could also think that only Landau local orders are thus sensitive (and fragile) to topological frustration: in~\cite{mari2021Nematic} we show that this is not the case and that nematic order can be destroyed by FBCs.

\section*{Acknowledgments}
We acknowledge support from the European Regional Development Fund -- the Competitiveness and Cohesion Operational Programme (KK.01.1.1.06 -- RBI TWIN SIN) and from the Croatian Science Foundation (HrZZ) Projects No. IP--2016--6--3347 and IP--2019--4--3321.
SMG and FF also acknowledge support from the QuantiXLie Center of Excellence, a project co--financed by the Croatian Government and European Union through the European Regional Development Fund -- the Competitiveness and Cohesion (Grant KK.01.1.1.01.0004).

\section*{ \Large Supplementary Information}

	This is the supplementary information for the paper Resilience of the topological phases to frustration. Here we diagonalize the Cluster-Ising chain and Kitaev chain, and compute the order parameter.

	\section{Cluster-Ising Chain}\label{appendix cluster}
	
	\subsection{Diagonalization}\label{diagonalization Cluster}
	
	The Cluster-Ising chain Hamiltonian
	\begin{equation}
	\label{Hamiltonian_1_supp}
	H=\cos\phi \sum\limits_{j=1}^{N} \sigma_j^x\sigma_{j+1}^x +\sin\phi\sum\limits_{j=1}^{N} \sigma_{j-1}^y \sigma_{j}^z \sigma_{j+1}^y
	\end{equation}
	in terms of Jordan-Wigner fermions
	\begin{equation}\label{suppJW}
	c_j=\Big(\bigotimes_{l=1}^{j-1}\sigma_l^z\Big)\frac{\sigma_j^x+\imath\sigma_j^y}{2}, \quad c_j^\dagger=\Big(\bigotimes_{l=1}^{j-1}\sigma_l^z\Big)\frac{\sigma_j^x-\imath\sigma_j^y}{2},
	\end{equation}
	reads
	\begin{equation}
	\begin{split}
	H=&-\cos\phi\Bigg[\sum_{j=1}^{N-1}(c_jc_{j+1}+c_jc_{j+1}^\dagger)-\Pi^z(c_Nc_1+c_Nc_1^\dagger)+\textrm{h.c.}\Bigg]\\
	&+\sin\phi\Bigg[\sum_{j=2}^{N-1}(c_{j-1}c_{j+1}-c_{j-1}c_{j+1}^\dagger)-\Pi^z(c_{N-1}c_1+c_{N}c_2-c_{N-1}c_1^\dagger-c_Nc_{2}^\dagger)+\textrm{h.c.}\Bigg].
	\end{split}
	\end{equation}
	
	Because of the presence of $\Pi^z$, the Hamiltonian is not quadratic in the fermions, but becomes such in each $\Pi^z$ parity sector. Namely, we can split the Hamiltonian as
	\begin{equation}
	\label{supp_Hamiltonian_2}
	H=\frac{1+\Pi^z}{2}H^+ \frac{1+\Pi^z}{2} + \frac{1-\Pi_z}{2}H^- \frac{1-\Pi_z}{2} \; ,
	\end{equation}
	where both $H^+$ and $H^-$ are quadratic. As such, they can be brought to a form of free fermions.
	
	This is achieved by first writing $H^\pm$ in terms of the Fourier transformed Jordan-Wigner fermions,
	\begin{equation}\label{fourier transformed suppJW fermions}
	b_q=\frac{1}{\sqrt{N}}\sum\limits_{j=1}^{N} c_j \ e^{-\imath qj} , \quad b_q^\dagger=\frac{1}{\sqrt{N}}\sum\limits_{j=1}^{N} c_j^\dagger \ e^{\imath qj}  ,
	\end{equation}
	for $q\in\Gamma^\pm$, where the two sets of momenta are given by $\Gamma^-=\{2\pi k/N \}$ and $\Gamma^+=\{2\pi (k+\frac{1}{2})/N \}$ with $k$ running over all integers between $0$ and $N-1$. Then the Bogoliubov rotation
	\begin{equation}\label{Bogoliubov particles}
	\begin{split}
	&a_q=\cos\theta_q \ b_q + \imath \sin\theta_q \ b_{-q}^\dagger, \quad q\neq0,\pi\\
	&a_{q}=b_q, \quad q=0,\pi 
	\end{split}
	\end{equation}
	with the Bogoliubov angle
	\begin{equation}\label{arctan}
	\theta_{q}=\arctan \frac{|\sin \phi + \cos \phi \ e^{\imath 3q}| +\cos\phi\cos q-\sin\phi \cos 2q}{\cos\phi \sin q-\sin\phi \sin2q}
	\end{equation}
	brings $H^\pm$ to a free fermionic form. We end up with
	\begin{equation}
	\label{supp_Hamiltonian_3}
	H^\pm=\sum\limits_{q\in\Gamma^\pm}^{} \varepsilon_q \left(a_q^\dagger a_q-\frac{1}{2}\right) ,
	\end{equation}
	where the energies are given by
	\begin{eqnarray}
	\varepsilon_q  & =2\sqrt{1 +\sin 2\phi \cos 3q }& \,\,\,\,\,\,\,\,\,\, \forall \,q\neq 0,\pi,\nonumber \\
	\varepsilon_0 & =2( \sin\phi+\cos\phi)& \,\,\,\,\,\,\,\,\,\, q=0\in \Gamma^- , \\ 
	\varepsilon_\pi& = 2(\sin\phi-\cos\phi) \nonumber &  \,\,\,\,\,\,\,\,\,\,q=\pi\in \Gamma^+.
	\end{eqnarray}
	
	The eigenstates of $H$ are formed starting from the vacuum states $\ket{0^\pm}$, which satisfy $a_q\ket{0^\pm}=0$ for $q\in\Gamma^\pm$, and applying Bogoliubov fermions creation operators, while taking care of the parity requirements in \eqref{supp_Hamiltonian_2}. The vacuum states are given by
	\begin{equation}\label{Bogoliubov vacuum}
	\ket{0^\pm}=\prod\limits_{0<q<\pi,\; q\in\Gamma^\pm} \big(\cos\theta_q-\imath\sin\theta_q \ b_q^\dagger b_{-q}^\dagger \big) \ket{0},
	\end{equation}
	where $\ket{0}$ is the vacuum for Jordan-Wigner fermions, satisfying $c_j\ket{0}=0$. In particular, $\ket{0}=\ket{\uparrow\uparrow...\uparrow}$ is the state of all spin up. The vacuum states $\ket{0^+}$ and $\ket{0^-}$ both have parity $\Pi^z=+1$ by construction. The parity requirements in \eqref{supp_Hamiltonian_2} imply that the eigenstates of $H$ belonging to $\Pi^z=-1$ sector are of the form $a_{q_1}^\dagger a_{q_2}^\dagger...a_{q_{m}}^\dagger\ket{0^-}$ with $q_i\in\Gamma^-$ and $m$ odd, while $\Pi^z=+1$ sector eigenstates are of the same form but with $q_i\in\Gamma^+$, $m$ even and the vacuum $\ket{0^+}$ used. The ground states are given explicitly in the main text.
	
	Let us also note one technical subtlety of the model. The Bogoliubov angle $\theta_q$, defined by \eqref{arctan} can become undefined for some modes $q\neq0,\pi$ also point-wise, by fine--tuning of the parameters $N$ and $\phi$. The Bogoliubov angle for these modes $\theta_q$ can be defined in the same way as for modes $q=0,\pi$ in the next section and the problem with them can be circumvented. These points do not have different expectation values of observables and can be neglected.

	\subsection{Majorana correlators}\label{sec cluster majorana}
	We are going to present the computation of two-point correlators of Majorana fermions
	\begin{equation}\label{Majorana_Dirac_supp}
	A_j=c_j^\dagger+c_j, \quad B_j=\imath(c_j^\dagger-c_j),
	\end{equation}
	in {the cluster phase in} some details, because essentially the same reasoning is valid also for the Kitaev chain. For this computation it is convenient to write the Hamiltonians $H^\pm$ in terms of positive energy fermions $d_q$, that we now define. For $q\neq0,\pi$ we put simply
	\begin{equation}\label{def d 1}
	d_q=a_q.
	\end{equation}
	For the modes $q=0,\pi$ the Bogoliubov angle \eqref{arctan} is undefined. We are going to define it also for these modes and use the analogoue of \eqref{Bogoliubov particles} to define $d_q$. First, we note that the Bogoliubov angle defined by \eqref{arctan} for $q\neq0,\pi$ satisfies
	\begin{equation} \label{exp 2 theta}
	e^{\imath 2\theta_{q}}=e^{-\imath 2\p} \frac{\sin\phi+\cos\phi \ e^{\imath 3\p}}{|\sin\phi+\cos\phi \ e^{\imath 3\p}|} \; .
	\end{equation}
	Although for modes $q=0,\pi$ the expression \eqref{arctan} is undefined, there is no problems with expression \eqref{exp 2 theta}. We exploit this property and define
	\begin{equation}
	\theta_q\equiv \frac{1}{2\imath}\log e^{\imath 2\theta_{q}}, \quad q=0,\pi,
	\end{equation}
	where by $e^{\imath2\theta_q}$ the expression on the right hand side of \eqref{exp 2 theta} is understood. Having $\theta_q$ we define, as in \eqref{Bogoliubov particles},
	\begin{equation}\label{def d 2}
	d_q=\cos\theta_q \ b_q + i \sin\theta_q \ b_{-q}^\dagger, \quad q=0,\pi.
	\end{equation}
	Since for $q=0,\pi$ we have
	\begin{equation}
	e^{\imath 2\theta_q}=\textrm{sgn}(\varepsilon_q)
	\end{equation}
	these definitions will result in the property that all fermions $d_q$ have positive energies, i.e. we can write
	\begin{equation}
	H^\pm=\sum\limits_{q\in\Gamma^\pm}^{}| \varepsilon_q| \left(d_q^\dagger d_q-\frac{1}{2}\right) .
	\end{equation}
	With these definitions the ground state of $H^-$ ($H^+$), let's denote it by $\ket{g,H^-}$ ($\ket{g,H^+}$) is the state that is annihilated by all $d_q$ for $q\in\Gamma^- \ (\Gamma^+)$, i.e. $d_q\ket{g,H^-}=0$.
	
	It is easy to see from the exact solution that the ground state $\ket{g}$ of the Cluster-Ising Hamiltonian $H$, coincides with $\ket{g,H^+}$ for $\phi\in(\pi/4,3\pi/4)$ and with $\ket{g,H^-}$ for $\phi\in(-3\pi/4,-\pi/4)$. We note that a typical effect of geometrical frustration \cite{mari2019frustration,mari2020frustration,Dong2016,Giampaolo2018}, which is not the case here, is that $\ket{g}$ does not coincide with either of them, because of the parity requirements in \eqref{supp_Hamiltonian_2}.
	
	Let us thus compute the Majorana correlation functions in the state $\ket{g,H^-}$, identical analysis can be made also for $\ket{g,H^+}$.
	
	From the definitions \eqref{def d 1} and \eqref{def d 2}, we obtain
	\begin{equation}
	b_q=\cos\theta_{q} d_q+\imath \sin\theta_q d_{-q}^\dagger.
	\end{equation}
	Now, using the definition \eqref{fourier transformed suppJW fermions} we get
	\begin{equation}
	c_j=\frac{1}{\sqrt{N}} \sum_{q\in\Gamma^-} (\cos\theta_{q} d_q-\imath \sin\theta_q  d_{-q}^\dagger)e^{\imath q j} ,
	\end{equation}
	from which we get easily the correlation functions
	\begin{align}
	\braket{c_jc_l}_{g,H^-}&=\frac{\imath}{2N}\sum\limits_{q\in\Gamma^-}\sin 2\theta_q e^{\imath q (j-l)}, \\\braket{c_jc_l^\dagger}_{g,H^-}&=\frac{1}{2N}\sum\limits_{q\in\Gamma^-}(1+\cos 2\theta_q) e^{\imath q (j-l)},
	\end{align}
	Finally, from the definition \eqref{Majorana_Dirac_supp} of Majorana fermions we get
	\begin{align}
	&\braket{A_jA_l}_{g,H^-}=\braket{B_jB_l}_{g,H^-}=\delta_{jl} \label{AA nf},\\
	&-\imath\braket{A_jB_l}_{g,H^-}=\frac{1}{N}\sum\limits_{q\in\Gamma^-}e^{i2\theta_{q}}e^{-i\p(j-l)}.  \label{AB nf}
	\end{align}
	The only difference in the ground state $\ket{g,H^+}$ is that the sum in \eqref{AB nf} would be over $\Gamma^+$ instead of $\Gamma^-$. In the limit of a large system the results are the same since the sum becomes an integral, {exponentially fast}. We have
	\begin{align}
	&\braket{A_jA_l}_{g,H^\pm}=\braket{B_jB_l}_{g,H^\pm}=\delta_{jl} ,\label{AA tl}\\
	&-\imath\braket{A_jB_l}_{g,H^\pm}\stackrel{N\to\infty}{\simeq}\int_0^{2\pi} e^{i2\theta_{q}}e^{-i\p(j-l)} \frac{dq}{2\pi} \label{AB tl}. 
	\end{align}
	
	{In the antiferromagnetic phase the ground state of the Cluster-Ising chain with FBCs is not anymore a vacuum state for positive energy fermions, i.e. the ground state coincides neither with the ground state of $H^+$ nor with the one of $H^-$. Instead, it corresponds to the vacuum state with one excitation on top of it. Correspondingly, the Majorana correlation functions acquire corrections of order $1/N$. For the ground state for $\phi\in(-\pi/4,0)$ when $N$ is divisible by $3$, presented in the main text, and given by
		\begin{equation}\label{supp gs three fold}
		\ket{g}=(u_1 a_0^\dagger+u_2 a_{\frac{2\pi}{3}}^\dagger+u_3 a_{-\frac{2\pi}{3}}^\dagger)\ket{0^-}.
		\end{equation}
		After some algebra we get
		\begin{equation}
		\begin{split}\label{AA superposition_correlator_supp}
		&\braket{A_jA_l}_g=\delta_{jl}-\frac{2\imath}{N}(|u_2|^2-|u_3|^2)  \sin\Big[\frac{2\pi}{3}(j-l)\Big]-\frac{2\imath}{N}\Big[(u_1^*u_2-u_3^*u_1)e^{\imath\frac{\pi}{3}(j+l-1)} +\textrm{c.c.}\Big] \sin\Big[\frac{\pi}{3}(j-l)\Big],\\
		&\braket{B_jB_l}_g=\delta_{jl}-\frac{2\imath}{N}(|u_2|^2-|u_3|^2)  \sin\Big[\frac{2\pi}{3}(j-l)\Big]-\frac{2\imath}{N}\Big[(u_1^*u_2-u_3^*u_1)e^{\imath\frac{\pi}{3}(j+l+1)} +\textrm{c.c.}\Big] \sin\Big[\frac{\pi}{3}(j-l)\Big],\\
		-\imath & \braket{A_jB_l}_g=\frac{1}{N}\sum\limits_{q\in\Gamma^-}e^{\imath 2\theta_{q}}e^{-\imath\p(j-l)} -\frac{2}{N}\Big\{ |u_1|^2+(|u_2|^2+|u_3|^2)\cos\Big[\frac{2\pi}{3}(j-l-1)\Big] \Big\} \\
		&-\frac{2}{N}\Big[(u_1^*u_2+u_3^*u_1) e^{\imath\frac{\pi}{3}(j+l)} +\textrm{c.c.} \Big]\cos\Big[\frac{\pi}{3}(j-l-1)\Big] - \frac{2}{N}\Big[u_2^*u_3 \ e^{-\imath\frac{2\pi}{3} (j+l)} +\textrm{c.c.}  \Big].
		\end{split}
		\end{equation}
		The ground state and the correlators when $N$ is not divisible by $3$ can be reproduced from these expressions by taking formally $u_2=u_3=0$.	
	}
	
	\subsection{Spin-correlation functions}
	{In this section we compute the spin-correlation functions $\braket{\sigma_1^x\sigma_{1+r}^x}_g$ in the ground state $\ket{g}$ in the antiferromagnetic phase of the model, given in the main text. We start from	the relation
		\begin{equation}\label{xx A B}
		\sigma_1^x\sigma_{1+r}^x=(-1)^r \prod_{j=1}^{r}(-iA_{j+1}B_j)
		\end{equation}
		and use the Wick theorem to reduce the spin-correlation functions to the pfaffian of the Majorana correlation matrix.}
	
	{Let us first discuss the applicability of the Wick theorem. When $N$ is not divisible by $3$, or when $N$ is divisible by $3$ and $u_j=1$ for some $j\in\{1,2,3\}$, it's easy to write the ground state as a vacuum state for some fermionic operators, so the Wick theorem can be applied. In a more general case when $N$ is divisible by $3$ it's a bit more complicated. If some coefficient $u_j$ is equal to zero then the same argument as in ref.\cite{mari2020frustration} can be given for the applicability. If all of them are non-zero we proceed in the following way. First, similarly to ref.\cite{mari2020frustration}, we define the fermions $\alpha_q$ by
		\begin{equation}
		\alpha_p=\frac{1}{(|u_2|^2+|u_3|^2)^{1/2}}\Big(u_2 a_p^\dagger +u_3 a_{-p}^\dagger \Big) , \quad \alpha_{-p}=\frac{1}{(|u_2|^2+|u_3|^2)^{1/2}}\Big(u_3 a_p -u_2 a_{-p} \Big),
		\end{equation}
		for $p\equiv 2\pi/3$, and by $\alpha_q=a_q$ for $q\neq p,-p$. Then we make another similar step and define the fermions $\beta_q$ by
		\begin{align*}
		\beta_0=\alpha_{-p}, \quad \beta_p=u_1a_0^\dagger+(|u_2|^2+|u_3|^2)^{1/2}\alpha_p, \quad \beta_{-p}=(|u_2|^2+|u_3|^2)^{1/2}a_0-u_1 \alpha_p^\dagger,
		\end{align*}
		and by $\beta_q=\alpha_q$ for $q\neq 0,p,-p$. Then the state \eqref{supp gs three fold} satisfies $\ket{g}=\beta_p\ket{0^-}$ and it's easy to see that it is annihilated by all $\beta_q$, i.e. we have $\beta_q\ket{g}=0$ for all $q\in\Gamma^-$. Thus, we have expressed the ground state as the vacuum for fermions $\beta_q$. Moreover, since Majorana fermions $A_j,B_j$ can be expressed as a linear combination of fermions $a_q,a_q^\dagger$ , they can also be expressed as a linear combination of $\beta_q, \beta_q^\dagger$. Therefore, Wick theorem can be applied.}
	
	{Applying the Wick theorem, we express the spin-correlation function as a pfaffian
		\begin{equation}\label{pfaffian}
		\braket{\sigma_1^x\sigma_{1+r}^x}_{g}=(-1)^{r+\lfloor\frac{r}{2}\rfloor} \ \textrm{pf}
		\begin{pmatrix}
		\mathbf{A} & \mathbf{C}\\
		-\mathbf{C}^\textrm{T} & -\mathbf{B}
		\end{pmatrix}.
		\end{equation}
		Here $\mathbf{A}$ and $\mathbf{B}$ are antisymmetric $r\times r$ matrices, defined by the elements $\mathbf{A}_{j,l}=\braket{A_{j+1}A_{l+1}}_{g}$ and $\mathbf{B}_{j,l}=\braket{B_{j}B_{l}}_{g}$ for $j<l$, while $\mathbf{C}$ is an $r\times r$ matrix with the elements $\mathbf{C}_{j,l}=-\imath \braket{A_{j+1} B_l}_{g}$ ($j$ and $l$ range from $1$ to $r$). In a more simple special case when the correlators $\braket{A_jA_l}_{g}$ and $\braket{B_jB_l}_{g}$ in the ground state $\ket{g}$ are zero for $j\neq l$, the spin correlations become simply the determinant
		\begin{equation}\label{special case of the pfaffian}
		\braket{\sigma_1^x\sigma_{1+r}^x}_{g}=(-1)^r\det \mathbf{C},
		\end{equation}
		as in ref.\cite{LIEB1961407}.}
	
	{Now let us compute the spin-correlation functions. When $N$ is not divisible by $3$ the correlators $\braket{A_jA_l}_{g}$ and $\braket{B_jB_l}_{g}$ are zero so we can use \eqref{special case of the pfaffian}. Approximating the sum in \eqref{AA superposition_correlator_supp} by integral we get
		\begin{equation}\label{C matrix element}
		\mathbf{C}_{j,l}\stackrel{N\to\infty}{\simeq}\int\limits_0^{2\pi}\frac{1+\tan\phi \ e^{-\imath 3 q}}{|1+\tan\phi \ e^{-\imath 3 q}|} e^{-\imath q(j-l)}\frac{dq}{2\pi} -\frac{2}{N},
		\end{equation}
		Without the second term, that stems from frustration, we would be able to apply strong Szeg\H{o} limit theorem \cite{TheTwo-DimensionalIsingModel,DeiftItsKrasovsky} to find the asymptotics of the Toeplitz determinant, and, therefore, of the spin-correlation functions. The second term is a correction, which can be understood as resulting from the part proportional to the delta function $\delta(q)$ in the symbol of the Toeplitz matrix $\mathbf{C}$. The asymptotics of such determinants has been studied in ref.\cite{MaricToeplitz}. The correction to the elements of the Toeplitz matrix results in a multiplicative correction to the determinant. Using Theorem 1 from there, in combination with the strong Szeg\H{o} limit theorem, we get
		\begin{equation}\label{supp spin correlations n 2}
		\braket{\sigma_1^x\sigma_{1+r}^x}_{g}\stackrel{r\to\infty}{\simeq}(-1)^r(1-\tan^2\phi)^{3/4} \bigg(1-\frac{2r}{N}\bigg).
		\end{equation}
		For a three-fold degenerate ground state when $N$ is divisible by $3$ the calculation is more complicated. Then we use directly \eqref{pfaffian} and resort to the numerical evaluation of pfaffians. However, we find that the result is the same, given by \eqref{supp spin correlations n 2}.}

	\subsection{Expectation value of the String operator}
	For completeness we also compute the ground state expectation value of the string operator
	\begin{equation}\label{string_supp}
	O(r)=\sigma_1^y\sigma_2^x \Big(\bigotimes_{j=3}^{r}\sigma_j^z \Big) \sigma_{r+1}^x\sigma_{r+2}^y.
	\end{equation}
	In terms of Majorana fermions \eqref{Majorana_Dirac_supp} the operator reads
	\begin{equation}
	O(r)=\prod_{j=1}^r(-\imath A_j B_{j+2}).
	\end{equation}
	Let us focus on the region $\phi\in(\pi/4,3\pi/4)$. Since the correlators $\braket{A_jA_l}_g$ and $\braket{B_jB_l}_g$ vanish for $j\neq l$, the expectation value of the string operator can be expressed, {using} Wick theorem, as a determinant
	\begin{equation}\label{O det}
	\braket{O(r)}_g=\det \mathbf{D},
	\end{equation}
	where $\mathbf{D}$ is an $r\times r$ correlation matrix with the elements
	\begin{equation}
	\mathbf{D}_{j,l}=-\imath\braket{A_jB_{l+2}}_g\stackrel{N\to\infty}{\simeq}\int_{0}^{2\pi}\frac{1+\cot \phi \ e^{\imath 3q}}{|1+\cot \phi \ e^{\imath 3q}|} e^{-\imath q(j-l)}\frac{dq}{2\pi}.
	\end{equation}
	For $\phi\in(-3\pi/4,-\pi/4)$ the only difference is that there is an additional factor $(-1)^r$ in front of the determinant in \eqref{O det}, because in this case $\sin\phi<0$ in \eqref{exp 2 theta}. The asymptotic behavior as $r\to\infty$ of the Toeplitz determinant $\det \mathbf{D}$ is obtained using the Strong Szeg\H{o} limit theorem \cite{TheTwo-DimensionalIsingModel,DeiftItsKrasovsky}. The result is \begin{equation}\label{string_supp order}
	\!\!\!\!\braket{O(r)}_g\!\!\stackrel{r\to\infty}{\simeq}\!\! \begin{cases}
	(1-\cot^2\phi)^{\frac{3}{4}} , & \phi\in(\frac{\pi}{4},\frac{3\pi}{4})\\
	(-1)^r (1-\cot^2\phi)^{\frac{3}{4}} , & \phi\in(-\frac{3\pi}{4},-\frac{\pi}{4})
	\end{cases}
	\end{equation}

	\section{Kitaev chain}\label{appendix Kitaev}
	
	The diagonalization of the Kitaev chain Hamiltonian
	\begin{equation}\label{Kitaev_Hamiltonian_supp}
	H=
	- \mu \sum_{j=1}^{N} \Big(c_j^\dagger c_j-\frac{1}{2}\Big)
	- \sum_{j=1}^{N} \left[
	w \: (c_j^\dagger c_{j+1}+\textrm{h.c.})
	-\Delta \: (c_jc_{j+1}+\textrm{h.c.}) \right]
	\end{equation}
	with periodic BC is very similar to the diagonalization of $H^-$ of the Cluster-Ising chain, discussed in section \ref{diagonalization Cluster}. The Hamiltonian is brought to a form of free fermions
	\begin{equation}
	H=\sum_{q\in\Gamma^-}\varepsilon_q \Big(a_q^\dagger a_q-\frac{1}{2}\Big),
	\end{equation}
	where $a_q$ are, again, Bogoliubov fermions, and the dispersion is now given by
	\begin{align}
	& \varepsilon_q=\sqrt{(4w\cos q+\mu)^2+4\Delta^2\sin^2 q} , \quad q\neq 0 ,\pi\\
	& \varepsilon_0=-2w-\mu,\\
	& \varepsilon_\pi=2w-\mu.
	\end{align}
	The Bogoliubov angle satisfies
	\begin{equation}
	\tan\theta_q=-\frac{|2w\cos q+\mu+2\Delta\sin q|+2w\cos q+\mu}{2\Delta\sin q}
	\end{equation}
	and
	\begin{equation}\label{exp 2 theta kitaev}
	e^{\imath 2\theta_q}=-\frac{2w\cos q+\mu+2\Delta\imath\sin q}{|2w\cos q+\mu+2\Delta\imath\sin q|}
	\end{equation}
	for $q\neq0,\pi$. Note that the mode $q=\pi$ does not exist with FBC, since $N$ is odd and momenta are quantized as integers.
	
	Since in the Kitaev chain we do not have parity restrictions like in \eqref{supp_Hamiltonian_2}, the ground state can always be written as a state annihilated by all positive energy fermions $d_q$, defined in section \eqref{sec cluster majorana}. This also implies that the Majorana correlation functions in the ground state are given by \eqref{AA tl} and \eqref{AB tl}, with $e^{\imath 2\theta_q}$ given by \eqref{exp 2 theta kitaev}. This is valid both for $N$ odd and $N$ even.

	\vskip .5cm
	
	\hrulefill 
	
	\vskip .5cm

\bibliographystyle{hunsrt}
\bibliography{bibliography}

\end{document}